\documentstyle[11pt,gh2001-asp,twoside,epsf]{article}
\begin{document}
\title{Secular Evolution in Barred Galaxies: Observations}
 \author{Michael R. Merrifield}
\affil{School of Physics \& Astronomy, University of Nottingham, Nottingham NG7 2RD, UK}

\begin{abstract}

This paper describes a framework for studying galaxy morphology,
particularly bar strength, in a quantitative manner, and presents
applications of this approach that reveal observational evidence for
secular evolution in bar morphology.  The distribution of bar strength
in galaxies is quite strongly bimodal, suggesting that barred and
unbarred systems are distinct entities, and that any evolution between
these two states must occur on a relatively rapid timescale.  Bars'
strengths appear to be correlated with their pattern speeds, implying
that these structures weaken as they start to slow, and disappear
entirely before the bars have slowed significantly.  There is also
tantalizing evidence that bars are rare beyond a redshift of $z \sim
0.7$, indicating that galaxies have only recently evolved to a point
where bars can readily form.

\end{abstract}

\section{Introduction}

I had originally intended to give an overview of the observational
evidence relating to the mechanisms by which bars evolve over time.
However, this overview has been expertly provided by a number of other
authors in these proceedings.  Bureau (2002), for example, gives a
thorough treatment of the buckling of bars to form boxy bulges, and
how this may lead to their dissolution.  Similarly, Das et al.\ (2002)
discuss the role that central mass concentrations, such as massive
black holes, may play in weakening and ultimately destroying bars.
Debattista (2002) describes the way that bars might be expected to
slow down through dynamical friction, while Gerssen (2002) presents
the observational evidence that not much slowing has occurred.  This
apparent contradiction suggests that the mass distribution in barred
galaxies cannot be very centrally concentrated, or that some other
mechanism destroys the bars before they have had the opportunity to
slow down.  Finally, Athanassoula (2002), through her sophisticated
N-body simulations, discusses much of the theory behind these
observations.  This wide sweep of articles leaves me very little to
say by way of background review.

I will therefore limit this paper to discussing how these
observations, and the accompanying theory, might be brought together
in a common quantitative framework, and to presenting a couple of
preliminary applications of this procedure.  Section~\ref{sec:quant}
discusses the quantification of bar strength, and how it can be
combined with other quantitative measures of morphology to define a
parameter space directly analogous to that used in the previous
qualitative classification of Hubble's tuning fork.
Section~\ref{sec:Omegap} describes an application of this
parameterization in studying the evolution of bar pattern speeds, and
Section~\ref{sec:HDF} looks at direct evidence for bar evolution in
the Hubble Deep Fields.  Finally, Section~\ref{sec:conc} speculates on
the next steps in this field.

\section{Quantifying Bar Strength and Morphology}
\label{sec:quant}

Historically, morphological features such as the presence or absence
of a bar in a galaxy have been assessed qualitatively by inspection of
photographic plates [e.g.\ Sandage \& Tammann 1981 (RSA), de
Vaucouleurs et al.\ 1991 (RC3)].  Although this approach has a strong
romantic appeal, it has a number of shortcomings in modern studies of
galaxy morphology.  First, such human inspection unavoidably has a
subjective element, which makes comparison between different analyses
rather difficult.  Second, projects such as the Sloan Digital Sky
Survey are now becoming so large that it is not practical to inspect
each galaxy image individually.  Third, such classifications do not
make full use of the quantitative data that is now available from CCD
images.  Although this under-utilization may not matter when
classifying high-quality images of nearby galaxies, it becomes vital
when analyzing small noisy images of very distant galaxies, where
every photon counts.  Finally, an objective quantitative measure of a
property such as bar strength is vital if one wishes to correlate
galaxies' ``barriness'' with their other properties to seek insight
into the way that bars form and evolve.

A number of methods have been advocated for measuring bar strength in
a quantitative manner.  The simplest involves just fitting isophotes
to a galaxy at different radii (e.g.\ Martin 1995).  The intrinsic
shape of the galaxy's isophotes can then be measured by deprojecting
it to face on.  This deprojection requires that one assume that the
galaxy can be treated as a thin disk, and that its inclination can be
estimated from the apparent shape of its outer isophotes, which are
taken to be intrinsically round.  A more robust variant on this
technique (Abraham et al.\ 1999) involves measuring the apparent
ellipticity as a function of radius by calculating the second moments
of the light distribution as a function of surface brightness
threshold.  Whichever approach is used, some measure of the bar
flattening, such as its intrinsic ellipticity or axis ratio, then
provides the measure of bar strength.  A more physically-motivated
approach has been advocated by Block et al.\ (2002), who seek to
measure the gravitational influence of a galaxy's non-axisymmetric
component relative to the axisymmetric forces.  Each of these measures
is optimized for different purposes: for example, the moment-based
technique is designed to deal with faint distant galaxies where the
signal-to-noise ratio is low, whereas the gravitational force method
is intended to tie the observations as closely as possible to the
underlying physics where higher quality data exist.  All the methods
have to make some assumption about the shapes of the galaxies surveyed
in order to derive their three-dimensional properties from
two-dimensional images.  In many cases, these assumptions will be
violated -- treating a galaxy as a thin disk when it contains a
significant bulge component is clearly incorrect.  However, even where
the assumptions are invalid, the technique will still provide a
measure of barriness.  As long as the measure is compared to the same
quantity derived from other samples that have been treated in the same
way, one can still draw meaningful conclusions from any differences
between the statistics from the various samples.

It seems very likely that bars cannot be fully described by any
such single parameter -- for example, is a small highly-distorted
central isophote a stronger or weaker bar than a much larger
slightly-distorted central feature?  Ultimately, one probably needs at
least two parameters to capture the essence of a bar, one describing
the degree to which it distorts the galaxy's isophotes, while the
other specifies the fraction of the galaxy's light that lies in the
bar.  However, even a single parameter does provide an objective and
repeatable measure of barriness which for the first time allows a
quantitative discussion of bar properties.

\begin{figure}
\plotone{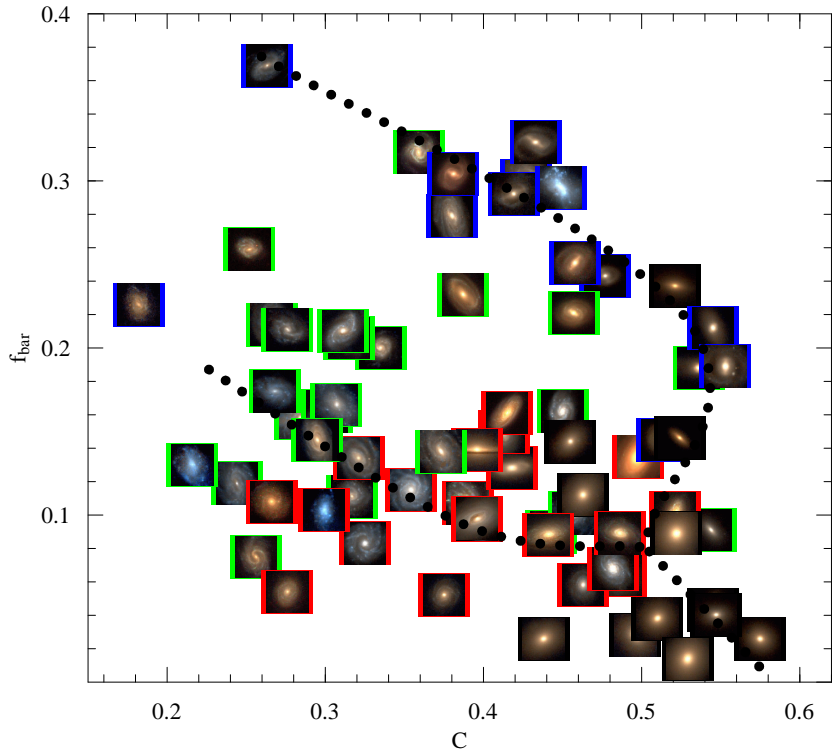}
\caption{The distribution of Frei Catalog galaxies (Frei et al.\ 1996)
in Hubble Space.  The possible locus of Hubble's Tuning Fork is
indicated.\label{fig:frei}}
\end{figure}

One simple use of such a measure of bar strength has been in an
investigation of whether Hubble's (1936) tuning fork classification of
galaxies represents a physical phenomenon, or whether it is just an
aesthetic idealization.  Hubble's scheme arranges the galaxies from
early type (big bulge, weak tightly-wound spiral structure) to late
type (small bulge, strong open spiral structure) along a horizontal
axis, bifurcating vertically into barred and unbarred galaxies.  Thus,
it can be thought of as a two-dimensional parameter space, with a
measure of lateness along the $x$ axis and a measure of barriness on
the $y$ axis.  Figure~\ref{fig:frei} shows a quantitative realization
of this ``Hubble Space,'' where Abraham et al.'s (1994) central
concentration index, $C$, is used to measure the position of a galaxy
along the Hubble sequence, and Abraham \& Merrifield's (2001) measure
of bar strength, $f_{\rm bar}$, delineates the $y$ axis.  The picture
is a little complicated here, partly because the high values of $C$
correspond to galaxies with large central concentrations of light,
which translate into the big bulges of early-type galaxies, so the
axis runs the opposite way to Hubble's traditional early-to-late
sequence.  Further, the parameters are somewhat correlated: the
appearance of a bar will be weakened in a galaxy with a strong central
bulge, whereas even a rather small bar shows up plainly in a late-type
galaxy with a weak bulge, so there is a general trend from top left to
bottom right in Fig.~\ref{fig:frei}.  However, despite these
complications, there does appear to be at least some sign of Hubble's
tuning fork in this picture, as the dotted line in the figure
indicates.  The barred galaxies, in particular, seem to form a
remarkably tight sequence, well separated from the main group of
unbarred systems.  A simple statistical test suggests that the deficit
of galaxies around values of $f_{\rm bar} \sim 0.2$, $C \sim 0.4$ is
significant at the 95\% level, although such {\it a posteriori}
calculations should not be over-interpreted.

If the gap between barred and unbarred galaxies is real, it has
important consequences for the secular evolution of bars.  It could
mean that barred and unbarred galaxies are fundamentally distinct
entities, with no evolution between the two types of system.  However,
these systems are remarkably similar in all their other properties.
For example, one might have thought that the presence or absence of a
bar could be related to how strongly the galaxy is dominated by dark
matter, which will alter the effectiveness of the classical bar
instability (Ostriker \& Peebles 1973).  The Tully-Fisher relation
essentially plots the correlation between luminous mass as quantified
by the absolute magnitude against total mass as quantified by the
rotation speed, so one might expect the relation to be different for
barred and unbarred galaxies.  However, the mean Tully Fisher relation
is identical for barred and unbarred systems (Debattista \& Sellwood
2000).  There is a somewhat larger scatter in the relation for barred
galaxies, but this can be attributed to the difficulty in measuring
the circular rotation speed of a barred galaxy where the stars and gas
do not follow circular orbits (Franx \& de Zeuuw 1992).  Given
that barred and unbarred galaxies seem otherwise identical, and that
there are the various credible mechanisms mentioned in the
Introduction by which bars might be destroyed, it seems unlikely that
barred and unbarred galaxies are fundamentally distinct.  More
plausibly, the absence of ``semi-barred'' galaxies can be taken to
show that galaxies evolve rapidly between the two extreme states --
the absence of intermediate galaxies is analogous to the Hertzsprung
gap in the color-magnitude diagram for stars, which represents a
region through which stars evolve rapidly.

This result conflicts with the RC3 classifications, which give a large
number of galaxies the ``SAB'' designation, indicating that they have
intermediate bar strength.  This conflict can be resolved by looking
at where the SAB galaxies are located in Fig.~\ref{fig:frei}.  They
are found to lie scattered around $C \sim 0.3$, $f_{\rm bar} \sim
0.15$; thus, they are the late-type galaxies on the unbarred branch of
Hubble's tuning fork.  Their values of $f_{\rm bar}$ are, indeed,
intermediate between those of the unbarred and strongly barred
galaxies, but this is simple because of the tilt in the tuning fork.
In essence, the SAB systems are unbarred galaxies that reveal minor
bar-like features at their centers; these fairly insignificant
features would have been hidden in galaxies with bigger bulges.

\begin{figure}
\plotone{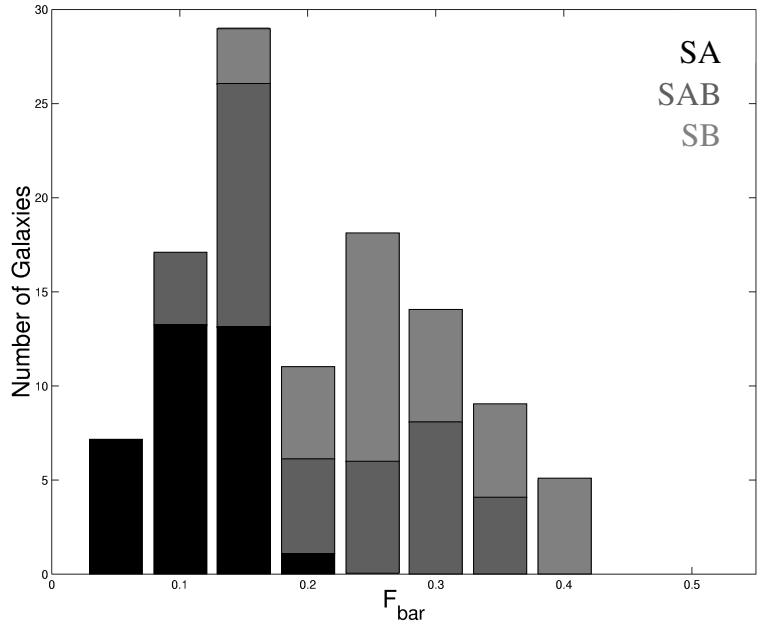}
\caption{The distribution of H-band bar strengths as measured in the
Ohio State University Bright Spiral Galaxy Survey.  The galaxies have
been separated into their SA, SAB and SB bar types.\label{fig:fbar}}
\end{figure}

One concern with the result presented in Fig.~\ref{fig:frei} is that
the morphologies have been derived from optical imaging, where dust
and localized star formation might disguise the underlying galaxy
morphologies (Block et al.\ 1999).  Further, it is not clear how the
galaxies in the Frei Catalog were selected.  It is therefore possible
that the bifurcation in Fig.~\ref{fig:frei} represents the natural
tendency to pick the extremes of a continuous distribution to
represent the archetypes of barred and unbarred systems.  To address
these concerns, Whyte et al.\ (2002) have analyzed H-band images
from the Ohio State University Bright Spiral Galaxy Survey.  These
galaxies form a relatively complete sub-sample from the RC3, so should
not have any bias toward artificial bimodality.  Nonetheless, as
Figure~\ref{fig:fbar} shows, there is still a significant dip in the
distribution of bar strengths at $f_{\rm bar} \sim 0.2$.  Since this
plot does not have the smearing effect of the tilt in the tuning fork
taken out, the true bimodality of the distribution is likely to be
even stronger.  Thus, the picture remains that any evolution from
barred to unbarred or vice-versa must happen on a rapid timescale
compared to the ages of the galaxies.

\section{Evolution in Morphology and Pattern Speed}
\label{sec:Omegap}

A second application of this quantitative approach to morphology is
provided by studies of bar pattern speeds.  A variety of techniques
have now been employed for measuring bar pattern speeds (e.g.\ Gerssen
2002).  The results all indicate that the bar patterns are rotating at
such a rate that their ends lie close to the point at which the
pattern rotates as rapidly as the galaxy, the co-rotation radius
$r_{\rm cr}$.  A bar can be conveniently parameterized by the
co-rotation radius divided by the length of its major axis, ${\cal R}
= r_{\rm cr}/a$, so the observations imply that ${\cal R} \sim 1$ for
all galaxies, which is the smallest value possible for any
self-consistent bar pattern.  This result is somewhat surprising
because, as described in Debattista (2002), a mechanism exists by which
bars evolve to ever lower pattern speeds until they essentially
grind to a halt, so one might expect to find bars with a wide range of
pattern speeds and values of ${\cal R}$ much greater than unity.

\begin{figure}
\plotone{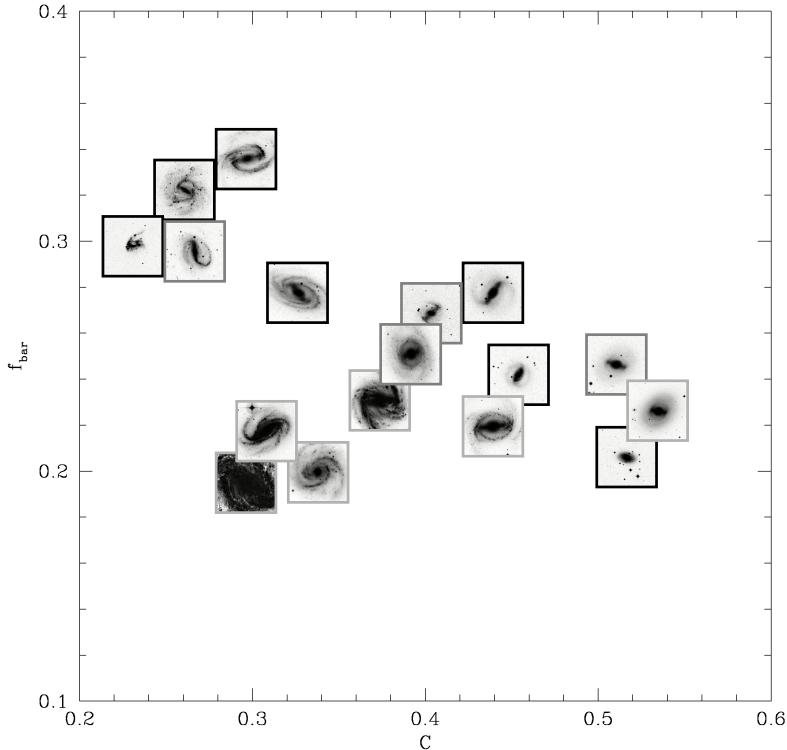}
\caption{Distribution in Hubble Space of barred galaxies with
measured pattern speeds.  The slower the pattern speed, the lighter the
border surrounding each image; however, even the slowest bars here
only have ${\cal R} \sim 1.8$.  Most of the pattern speeds come from the
compilation in Elmegreen (1996).\label{fig:omegap}}
\end{figure}

Some clue as to what is going on can be gleaned by looking at the
distribution in Hubble Space of galaxies with measured bar pattern
speed.  As Figure~\ref{fig:omegap} shows, these galaxies describe the
expected fairly tight sequence which matches that already derived for
barred galaxies.  However, there are indications that the pattern
speeds vary systematically across the width of the sequence, with
somewhat slower pattern speeds being associated with weaker bars.
This result would suggest that barred galaxies follow evolutionary
tracks that lead them downward in this figure as their bars start to
slow.  The absence of any slower bars suggest that any such galaxies
have entered the gap region where systems evolve rapidly via some
other mechanism that destroys the bar.

Clearly, however, larger samples of bars with measured pattern speeds
are required if this result is to be confirmed.  It will also be
highly instructive to calculate the Hubble Space parameters for the
bars that form and slow down in N-body simulations (Debattista 2002)
in order to follow their evolutionary tracks in this space, to see how
well they match up to the observations.

\section{Direct Observations of Bar Evolution}
\label{sec:HDF}

With the more robust measures of bar strength, such as the moment-based
techniques, it is possible to characterize this property even in small
faint galaxies, such as those imaged in the Hubble Deep Fields (HDFs).
This possibility is particularly intriguing as it allows one to
investigate the secular evolution of barred galaxies over a
cosmological timescale by examining the properties of bars in the
population as a function of redshift.

\begin{figure}
\plotone{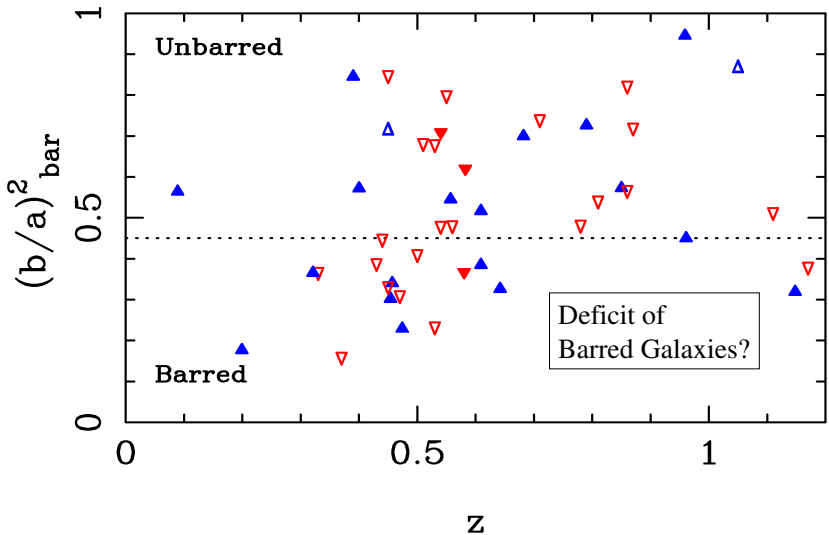}
\caption{Bar strength as a function of redshift for galaxies in the
Hubble Deep Fields.  Upward pointing triangle are from HDF-N, while
downward pointing triangles are from HDF-S.  Filled symbols are for
galaxies with spectroscopic redshifts, while open symbols rely on
photometric redshifts.  The dotted line indicates the value of the
central intrinsic axis ratio which, in nearby galaxies, separates
systems classified as barred from those deemed unbarred.\label{fig:hdf}}
\end{figure}

An analysis of the HDFs, summarized in Fig.~\ref{fig:hdf}, led Abraham
et al.\ (1999) to conclude that there has been strong evolution in the
barred galaxy population to the extent that there are almost no
barred systems beyond a redshift of $z \sim 0.7$.  This result
quantified the qualitative impression that van den Bergh et al.\
(1996) had obtained from visual inspection of the first HDF data.
This observation lends itself to the simple physical interpretation
that at these early epochs disks were either too dynamically hot or
too low in mass to undergo the conventional bar instability.

There are, however, several possible snags with this analysis.  The
first issue is bandshifting: all the data analyzed came from I-band
images, but at significant redshifts the observed I-band light will
have been emitted further to the blue, so one is not comparing like
with like when looking at galaxies at different redshifts.  However,
the effects here are fairly modest: at the redshifts where there is an
absence of bars, the I-band images correspond to rest-frame B-band
emission.  Since B-band images of nearby galaxies reveal plenty of
barred systems, the absence of bars beyond $z \sim 0.7$ cannot simply
be attributed to the observations being made in waveband where bars
are not detectable.  

The second problem is that the more distant galaxies tend to have
lower signal-to-noise ratios, which could prevent their bars being
detected.  One subtler effect is that cosmological dimming may reduce
the surface brightness of a galaxy's disk to an undetectable level, so
a face-on barred galaxy at high redshift may be misidentified as an
edge-on system.  Since edge-on systems cannot have their bar
strengths measured, they are excluded from the analysis, which might
explain where the barred systems have gone.  However, Abraham et al.'s
(1999) experiments involving artificially redshifting and degrading
the images of the closer galaxies showed that their bars would have
remained detectable even at the highest redshifts observed, and that
the disks remain bright enough for there to be no confusion over
inclination.  

It is still possible that barred galaxies exist at high redshifts: if,
for example, these systems have much weaker disks than those in the
nearby Universe, then they could have been misidentified as edge-on
systems.  One is, however, forced to the conclusion that if there are
barred galaxies beyond $z \sim 0.7$, then they are sufficiently
different in structure from those in the nearby Universe that they
cannot be detected in the HDF data.

\section{Conclusions and Future Work}
\label{sec:conc}

With the advent of largescale CCD surveys of galaxies, both nearby and
in the more distant Universe, the study of galaxy morphology has
matured from the qualitative to the quantitative.  Robust techniques
have been developed for automatically measuring properties of galaxies
such as their bulge-to-disk ratios and bar strengths.  Perhaps
unsurprisingly, these quantitative measures correspond quite closely
to their qualitative predecessors, to the extent that one can
reproduce Hubble's tuning fork classification.  The bifurcation
between barred and unbarred galaxies has significant implications for
investigations of secular evolution in bars, since it implies that any
evolution between these two end states must occur on a timescale that
is short compared to the galaxies' cosmological lifetimes.  Further
evidence for evolution in bar strength on timescales that are less
than the cosmological have come from studies of faint galaxies in the
HDFs, which indicate that bar formation is a relatively recent
phenomenon.  

An important next step will be to use this framework to compare the
observations with the predictions of N-body simulations.  It is now
possible to calculate morphological parameters to follow the
properties of N-body simulations, such as those described by
Athanassoula (2002), as the systems undergo secular evolution.
Further, cosmological simulations like those presented by Navarro
(2002) are rapidly advancing to the point where they can reliably
follow the formation and evolution of individual disk galaxies, so we
can also follow the variations in morphological parameters in a
cosmological framework.  Through such analysis, galaxy
observations such as those presented here can be directly tied to N-body
simulations to test our understanding of the formation and evolution
of these systems, just as the observed sequences of stars in the
color-magnitude diagram can be combined with stellar evolutionary
tracks in order to understand how stars form and evolve.

One other area where there is still much room for analysis lies in the
morphological description of bars.  As described in
Section~\ref{sec:quant}, it seems unlikely that the essence of a bar
can be captured in a single parameter, but it is by no means clear
what combination of parameters does provide the optimal description,
nor what the physical differences are that give rise to the range of
parameters observed.  A fuller description of morphological features
such as bars will enable us to bring more sophistication to our
exploration of the mechanisms by which these features evolve, and
hopefully answer many of the outstanding questions.


\begin{references}

\reference Abraham, R.G. \& Merrifield, M.R. 2000, AJ, 120, 2835

\reference Abraham, R.G., Merrifield, M.R., Ellis, R.S., Tanvir,
N.R. \& Brinchmann, J. 1999, MNRAS, 308, 569

\reference Abraham, R.G., Valdes, F., Yee, H.K.C. \& van den Bergh, S. 
1994, ApJ, 432, 75

\reference Athanassoula, L. 2002, this volume

\reference Block, D.L., Puerari, I., Frogel, J.A., Eskridge, P.B.,
Stockton, A. \& Fuchs, B. 1999, Ap\&SS, 269, 423

\reference Block, D.L., Puerari, I., Knapen, J.H., Elmegreen, B.G.,
Buta, R., Stedman, S., Elmegreen, D.M. 2002, this volume

\reference Bureau, M. 2002, this volume

\reference Das, M., Teuben, P.J., Vogel, S.N., Regan, M.W., Sheth,
K. \& Harris, A. 2002, this volume

\reference Debattista, V. 2002, this volume

\reference Debattista, V. \& Sellwood, J. 2000, ApJ, 543, 704

\reference de Vaucouleurs, G., de Vaucouleurs, A., Corwin, H.G., Buta, R.J.,
Paturel, G. \& Fouque, P. 1991, Third Reference Catalog of Bright
Galaxies (New York: Springer-Verlag)

\reference Elmegreen, B. 1996, in ASP Conf.\ Ser.\ Vol.\ 91, Barred
Galaxies, ed.\ R. Buta, D.A. Crocker \& B.G. Elmegreen (San Francisco:
ASP), 197

\reference Franx, M. \& de Zeuuw, T. 1992, ApJ, 392, L47

\reference Frei, Z., Guhathakurta, P., Gunn, J.E. \& Tyson, J.A. 1996,
AJ, 111, 174

\reference Gerssen, J. 2002, this volume

\reference Hubble, E.P. 1936, The Realm of the Nebulae (New Haven:
Yale University Press)

\reference Martin, P. 1995, AJ, 109, 2428

\reference Navarro, J. 2002, this volume

\reference Ostriker, J.P. \& Peebles, P.J.E. 1973, ApJ, 186, 467

\reference Sandage, A. \& Tammann, G.A. 1981, Revised Shapley-Ames Catalog
of Bright Galaxies (Washington: Carnegie Institute)

\reference van den Bergh S., Abraham R.G., Ellis R.S., Tanvir N.R. \&
Santiago B.X. 1996, AJ, 112, 359

\reference Whyte, L.F., Abraham, R.G., Merrifield, M.R., Eskridge,
P.B. \& Frogel, J.A. 2002, in preparation

\end{references}
\end{document}